\documentstyle[12pt,epsfig]{article}        
\newlength{\dinwidth}                       
\newlength{\dinmargin}                      
\def\lsim{\mathrel{\rlap{\lower4pt\hbox{\hskip1pt$\sim$}}
    \raise1pt\hbox{$<$}}}         
\def\gsim{\mathrel{\rlap{\lower4pt\hbox{\hskip1pt$\sim$}}
    \raise1pt\hbox{$>$}}}         

\def\be{\begin{equation}}
\def\ee{\end{equation}}
\def\bq{\begin{eqnarray}}
\def\eq{\end{eqnarray}}

\begin{document}
\vspace*{1cm}
\begin{center}  \begin{Large} \begin{bf}
The Charm--Strange Contribution \\ to Charged--Current DIS 
Structure Functions \\
  \end{bf}  \end{Large}
  \vspace*{5mm}
  \begin{large}
V.~Barone$^a$, U.~D'Alesio$^{a}$
 and 
M.~Genovese$^{b,}$\footnote{Supported by EU Contract ERBFMBICT
950427.}
\\ 
  \end{large}
\end{center}
$^a$ Universit{\`a} di Torino and INFN, 10125 Torino, Italy \\
$^b$ Universit\'e Joseph Fourier and IN2P3-CNRS, 38026 Grenoble, 
France \\
\begin{quotation}
\noindent
{\bf Abstract:}
We review the present theoretical knowledge of the charm--strange
contribution to charged--current DIS structure functions. In particular, 
the
uncertainties arising from the choice of the 
factorization scale, of the massive QCD scheme,  
and of the parton fit
are discussed. 
\end{quotation}

\section{Cross Sections}

Charm production in charged--current (CC) 
deep inelastic scattering (DIS) is the best 
way to obtain information on the strange sea 
density \cite{B}, 
which is at present the most poorly known among the light--quark
distributions. The strange distribution 
can also be obtained by properly combining 
fully inclusive CC cross sections.
CC reactions are thus necessary in order to
determine the strange sector and, in general, to reconstruct the whole 
flavor systematics.

The relevant subprocesses for charm excitation in CC DIS are
(we consider only Cabibbo unsuppressed diagrams): 

{\it i)}
the 
${\cal O}(\alpha_s^0)$ 
direct transition 
$W^{+}s \rightarrow c$; 

{\it ii)}
the ${\cal O}(\alpha_s^1)$ $W$-gluon and $W$-quark fusion 
processes $W^{+}g \rightarrow 
\bar s c\,, \;\;
W^{+}s \rightarrow g c.$ 

The CC DIS cross section reads 
\be
\frac{{\rm d}^2 \sigma^{\pm}}{{\rm d}x {\rm d}y}
= \frac{G^2 s}{4\pi (1 + \frac{Q^2}{M_W^2})^2}
\, \left [ x y^2 F_1 + (1-y - \frac{m_N^2 xy}{s}) F_2 \mp
(y - \frac{y^2}{2}) x F_3 \right ] \,,
\label{4}
\ee
where the subscript $+$ ($-$) denotes the reaction $e^{+}N 
\rightarrow \bar \nu X$ ($e^{-}N 
\rightarrow  \nu X$). 
If we restrict ourselves to charm excitation we get
at order $\alpha_s^0$ (for electron scattering)
\be
\frac{{\rm d}^2
\sigma_{cs}}{{\rm d}\xi {\rm d}y}
= \frac{G^2 s}{4\pi (1 + \frac{Q^2}{M_W^2})^2}
\, 2 \xi \, \bar s(\xi,Q^2) \vert V_{cs} \vert^2 
\, [ (1- y)^2 + \frac{m_c^2}{s \xi} \, (1-y) ]
\,\,,
\label{5}
\ee
where the slow rescaling variable
$\xi = x \, (1 + m_c^2/Q^2 )$
accounts for the finite mass of the charmed quark in the $W s 
\rightarrow c$ transition. The cross section for positron scattering
is obtained by the replacement $\bar s \rightarrow  s$.  
The cross section (\ref{5}) 
provides a direct measure of the strange sea density. 

However, at order $\alpha_s$ (often referred to, somehow improperly, as 
the next--to--leading order) formula (\ref{5}) does not 
hold any longer because of the more 
complicated relation between structure functions 
and parton densities. 
The most important ${\cal O}(\alpha_s)$ contribution 
comes from the vector-boson--gluon fusion term which 
incorporates important dynamical effects
(quark--mass threshold effects 
and large longitudinal contributions 
due to the non conservation of 
weak currents \cite{BGNPZ2}).

\section{Theoretical Uncertainties}

The QCD analysis of the charm--strange 
structure function at order $\alpha_s$ is affected by theoretical
uncertainties which have two sources:  
{\it i)} the choice of the massive QCD scheme; {\it ii)} the 
arbitrariness of the factorization scale. Besides these,  
there is a further uncertainty coming from the choice of 
the parton fit among those available on the market.

The most commonly used ${\cal O}(\alpha_s)$  schemes for massive 
quarks are the Fixed Flavor
Scheme (FFS) \cite{GR} and the Variable Flavor Scheme (VFS) \cite{VFS}. 

In the FFS, charm is treated as a  heavy quark, 
in an absolute sense. There is no charm 
excitation term in the structure functions and 
the number of active flavors is $N_f = 3$. 
The collinear divergence
$\log (Q^2/m_s^2)$ in the gluon fusion term is regularized 
at a scale $\mu^2$. For the $cs$ contribution to the 
$F_2$ structure function one explicitly has
\be
{\rm FFS:}\;\;\;\;\;
F_{2}^{cs}(x,Q^2) = 2\, \xi \, s(\xi, \mu^2) +
\frac{\alpha_s(\mu^2)}{2 \pi}  \,
\int_{\xi}^{1}
\frac{{\rm d}z}{z} \,\, 2 \xi\, [C_2^g(z, \mu^2)\,  
g (\xi/z, \mu^2) +  C_2^q(z, \mu^2)\,  
s (\xi/z, \mu^2) ]\,,
\label{ffs}
\ee
where the Wilson coefficients $C_2^g$ and $C_2^q$ 
can be found in the literature \cite{G}. 

In the VFS, charm is a ``heavy'' flavor for $\mu^2 < m_c^2$, 
and a 
partonic constituent of the nucleon for $\mu^2 > m_c^2$. 
Both the strange and 
the charm excitation terms appear and hence there are two subtractions
in the Wilson coefficient, corresponding to the two 
singularities in the limits $m_s \rightarrow 0$ and 
$m_c^2 / Q^2 \rightarrow 0$. The explicit expression is  
\bq
{\rm VFS:}\;\;\;\;\;
F_{2}^{cs}(x,Q^2) &=& 2\, \xi \, [ s(\xi, \mu^2) + c(x, \mu^2) ]
\nonumber \\
&+& \frac{\alpha_s(\mu^2)}{2 \pi}  \,
\int_{\xi}^{1}
\frac{{\rm d}z}{z} \,\, 2 \xi\, [ \hat C_2^g(z, \mu^2)\,  
g (\xi/z, \mu^2) +  \hat C_2^q(z, \mu^2)\,  
s (\xi/z, \mu^2) ]\,,
\label{vfs}
\eq
where $\hat C_{2}^{g,q}$ denotes the doubly subtracted
massive Wilson coefficients.

The factorization scale $\mu^2$ is arbitrary and only an educated 
guess can be made on it. It is clear that a knowledge 
of the CC structure functions at order $\alpha_s^2$, still 
lacking at present, would allow testing the perturbative stability
of the various choices.

\section{Results}

We estimate now the theoretical uncertainties on $F_2^{cs}$ and 
on the $cs$ contribution to the DIS cross section. 

\begin{figure}[htb] 
\mbox{\epsfig{file=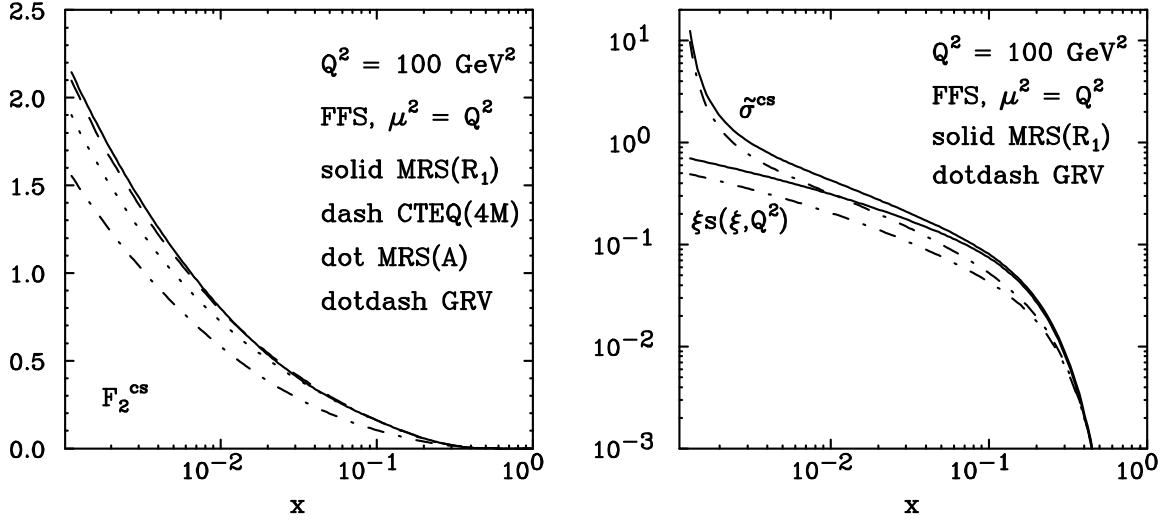,width=0.8\textwidth}}
\caption{\label{FIT}
{\em $F_2^{cs}$ and $ \tilde \sigma^{cs}$
in the FFS 
for various parton fits.}} 
\end{figure}

In Fig.~\ref{FIT} 
we show the results of the
calculation of $F_2^{cs}$ and of
\be
\tilde \sigma^{cs} 
\equiv K\,
 \frac{d \sigma^{cs}}{dx \, dy}\,\,, \hspace{35pt}
K^{-1} =
\frac{G^2 s}{2 \pi \, (1 + Q^2/M_W^2)^2} \,
\vert V_{cs} \vert^2 
\, [ (1- y)^2 + \frac{m_c^2}{s \xi} \, (1-y) ] \, ,
\label{sigmacs} 
\ee
at $Q^2 = 100$ GeV$^2$, 
for various NLO parton fits: 
MRS(A) \cite{MRS1}, MRS(R$_1$) \cite{MRS2}, CTEQ(4M) \cite{CTEQ}, 
GRV \cite{GRV}. The scheme used is the 
FFS and the factorization scale is taken to be  $\mu^2 = Q^2$. 
In the box on the right we also display $\xi  s(\xi, Q^2)$, 
that is what 
$\tilde \sigma^{cs}$ reduces to at order $\alpha_s^0$. 
Note that the CTEQ(4M) and MRS(R$_1$) curves nearly coincide 
whereas there is a non negligible difference between 
the two MRS fits and a larger discrepancy between 
MRS(R$_1$) and GRV. The global uncertainty due to 
the choice of the fit amounts to $\sim 30 \%$ both for 
$F_2^{cs}$ and for $\tilde \sigma^{cs}$.

\begin{figure}[htb] 
\mbox{\epsfig{file=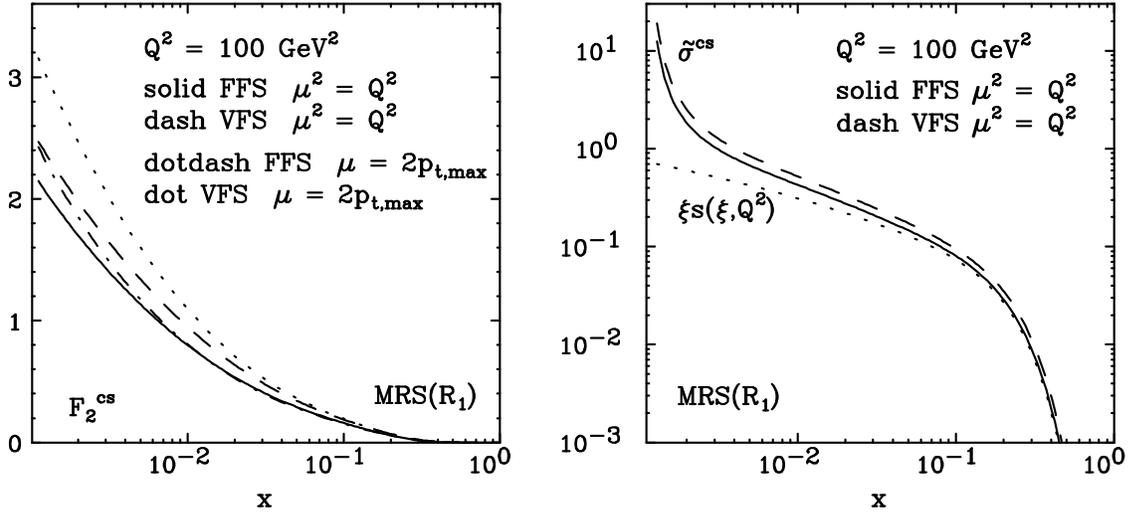,width=0.8\textwidth}}
\caption{\label{SCH}
{\em $F_2^{cs}$ and $\tilde \sigma^{cs}$
in the two schemes (FFS and VFS) and 
with two choices of the factorization scale.}} 
\end{figure}

In Fig.~\ref{SCH}, considering now only the MRS(R$_1$) parametrization, 
we illustrate the scheme dependence for two different factorization 
scales: $\mu^2 = Q^2$ and $\mu = 2 p_{t,max}$, where $p_t$ 
is the transverse momentum of the produced charmed quark. 
The difference between the FFS and the VFS results at 
$Q^2 = 100$ GeV$^2$ is again up
to $\sim 30-40 \%$ 
(attaining the largest value for $\mu = 2 p_{t,max}$). 

\begin{figure}[htb] 
\mbox{\epsfig{file=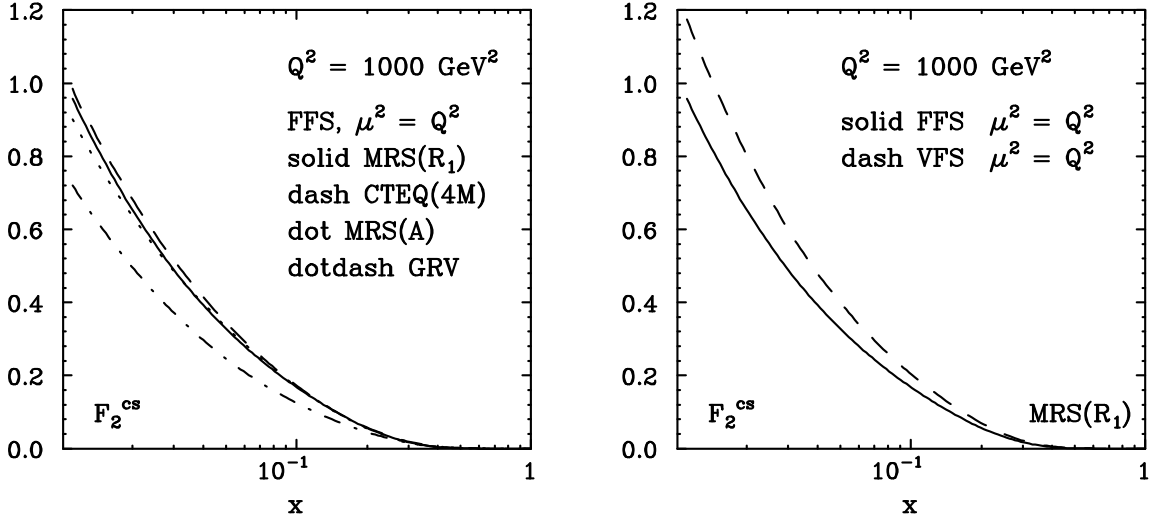,width=0.8\textwidth}}
\caption{\label{F21000}
{\em Fit and scheme dependence of $F_2^{cs}$ at $Q^2=1000$ GeV$^2$.}} 
\end{figure}

In Fig.~\ref{F21000} we show the situation at a higher 
physical scale, $Q^2=1000$ GeV$^2$. Notice that the difference
between the two schemes is still relatively large whereas 
the choice $\mu = 2p_{t,max}$ gives curves (not displayed) which 
are indistinguishable from those corresponding to $\mu^2 = Q^2$.

\begin{figure}[htb] 
\mbox{\epsfig{file=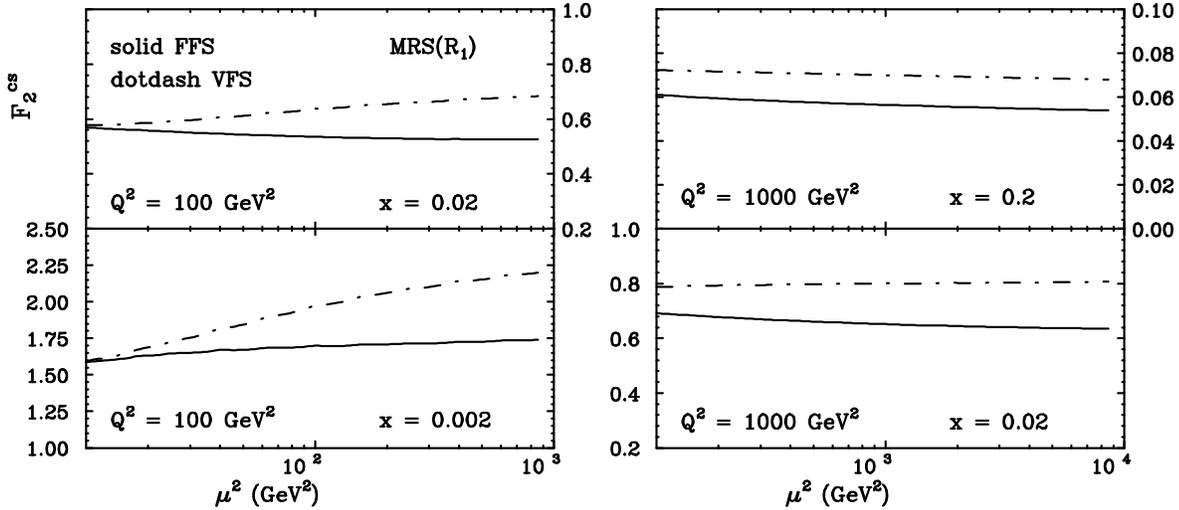,width=0.8\textwidth}}
\caption{\label{F2FAC}
{\em Dependence of $F_2^{cs}$ on the factorization scale in the two 
massive QCD schemes.}} 
\end{figure}

In Fig.~\ref{F2FAC} we illustrate the dependence on the 
factorization scale. 
The parton fit used is the MRS(R$_1$) 
and the results of 
both schemes are presented for 
some kinematically accessible $(x,Q^2)$ bins. 
It is clearly seen that when $Q^2$ is not very
large the VFS is more unstable than the FFS. 
At very high $Q^2$ both the VFS and the FFS 
are not very sensitive to the factorization scale.

It is important for HERA to know also the expected uncertainties
on the charm--strange contribution to the total cross section. These 
are presented in Table~1. It is interesting to notice that the main 
difference ($\sim$ 40 \%) arises from the choice of the scheme.

\begin{table}[htb] \label{Table1}
\centering
\begin{tabular}{||c|c|c|c||}	\hline
 $\sigma^{cs}_{tot}\; (pb)$ 
               & FFS/MRS(R$_1$) & FFS/GRV & VFS/MRS(R$_1$)\\ \cline{1-4}
 $\mu^2=Q^2$      & 4.66        &  4.02   &   6.79      \\ \cline{1-4}
 $\mu=2p_{t,max}$ & 4.20        &  3.57   &   6.80     \\ \hline
\end{tabular}
\caption{{\em Total charm-strange cross-section for different parton 
fits, factorization scales and schemes in the kinematic region 
$ Q^2 > 200$ GeV$^2$  and $ x > 0.006 $ 
( $ \protect\sqrt{s} \protect\simeq 300 $ GeV).}}
\end{table}

\section{Conclusions}

The overall  theoretical uncertainty on the charm--strange 
contribution to the
charged--current structure functions is relatively large, 
being at least of order of 30 \%
in the typically accessible $(x,Q^2)$ region. 
At not very high $Q^2$ the Fixed Flavor Scheme
turns out to be preferable due to its greater stability. 
An order $\alpha_s^2$ calculation 
is necessary to settle the problem 
of the scheme and factorization scale dependence. 
For the sake of consistency 
and for a safer analysis it would be important to 
use massive QCD evolution in the heavy--quark sector of the 
global fits. Possible HERA data on the charm--strange
structure functions 
at large $Q^2$ would certainly be of the greatest utility.

\end{document}